\documentclass[11pt]{article}

\usepackage{epsfig}
\usepackage{a4}

\newif\iffigframe

\newif\ifbigfig

\newcommand{\figbox}[3]{\hbox to#1\bgroup
  \dimen0=1bp \dimen1=#1\relax
  \def\a##1 ##2 ##3 ##4 ##5\\{\if!##1!\a##2 ##3 ##4 ##5 .\\\else
    \dimen3=##3\dimen0 \advance\dimen3 -##1\dimen0
    \dimen4=##4\dimen0 \advance\dimen4 -##2\dimen0
    \dimen5=\dimen4 \divide\dimen5 \dimen3
    \dimen2=\dimen1 \multiply\dimen2 \dimen5
    \multiply\dimen5 \dimen3 \advance\dimen4 -\dimen5
    \dimen5=\dimen1
    \loop \advance\dimen4 \dimen4 \divide\dimen5 2
    \ifnum\dimen5>0 \ifnum\dimen4<\dimen3 \else
      \advance\dimen4 -\dimen3 \advance\dimen2 \dimen5 \fi
    \repeat
    \dimen5=10\dimen1 \divide\dimen5 \dimen0
    \includegraphics{#3.eps}%
    \iffigframe \vrule\hss \else \hfil \fi
    \vbox to\dimen2\bgroup
      \iffigframe \hrule width\dimen1\vss \hrule \else \vfil \fi
      \egroup
    \iffigframe \vrule\hss \fi
    \egroup\fi}%
  \a#2 . . . .\\}

\newcounter{subequation}[equation]

\makeatletter
\expandafter\let\expandafter\reset@font\csname reset@font\endcsname
\newenvironment{subeqnarray}
  {\arraycolsep1pt
    \def\@eqnnum\stepcounter##1{\stepcounter{subequation}{\reset@font\rm
      (\theequation\alph{subequation})}}\eqnarray}%
  {\endeqnarray\stepcounter{equation}}
\makeatother

\newcounter{statement}
\newenvironment{statement}[4]
  {\par\refstepcounter{statement}
    \noindent#1#2 \arabic{statement} #4\unskip: #3}{\par\vspace{2mm}}

\newenvironment{statement*}[4]
  {\par\noindent#1#2 #4\unskip: #3}{\par\vspace{2mm}}

\newcommand{\lam}{|\Lambda|}

\begin{document}

\hbox to\hsize{%
  \vbox{%
        }\hfil
  \vbox{%
        \hbox{MPI-PhT/2003-34}%
        \hbox{July 2, 2003}%
        \hbox{gr-qc/0307029}%
        }}

\vspace{1cm}
\begin{center}
\LARGE\bf
Non-Abelian gravitating solitons with
negative cosmological constant

\vskip5mm
\large Peter Breitenlohner,
Dieter Maison\\
\vspace{3mm}
\small\sl
Max-Planck-Institut f\"ur Physik\\
--- Werner Heisenberg Institut ---\\
F\"ohringer Ring 6\\
D-80805 Munich, Germany\\
\vspace{3mm}
{\normalsize and}\\
\large\bf
George Lavrelashvili\\
\vspace{3mm}
\small\sl
Department of Theoretical Physics\\
A.Razmadze Mathematical Institute\\
GE-0193 Tbilisi, Georgia\\

\end{center}
\vspace{10mm}
\begingroup \addtolength{\leftskip}{1cm} \addtolength{\rightskip}{1cm}
\subsection*{Abstract}

Static, spherically symmetric solutions with regular origin are
investigated of the Einstein-Yang-Mills theory with a negative
cosmological constant $\Lambda$. A combination of numerical and
analytical methods leads to a clear picture of the `moduli space'
of the solutions. Some issues discussed in the existing literature
on the subject are reconsidered and clarified. In particular the
stability of the asymptotically AdS solutions is studied. Like for
the Bartnik-McKinnon (BK) solutions obtained  for $\Lambda=0$
there are two different types of instabilities -- `topological'
and `gravitational'. Regions with any number of these
instabilities are identified in the moduli space. While for BK
solutions there is always a non-vanishing equal number of
instabilities of both types, this degeneracy is lifted and there
exist stable solutions, genuine sphalerons with exactly one
unstable mode and so on. The boundaries of these regions are
determined.

\endgroup

\section{Introduction}

The last decade has witnessed an intensive study of soliton and black hole
solutions of the Einstein-Yang-Mills (EYM) theory leading to many unexpected
results (see \cite{volkov} for a review).
In particular it turned out that the strong uniqueness theorems for such
solutions valid for gravitating Abelian gauge fields (Einstein-Maxwell
theory) cease to hold in the non-Abelian
case. Also the stability properties of spherically symmetric solutions
change, partly due to the breakdown of the Birkhoff theorem in the non-Abelian
theory.

Among the many possible extensions of the EYM system studied in
the literature the simplest one is obtained by adding a
cosmological constant $\Lambda$ \cite{tmt95,vslhb96}. As to be
expected this changes the global behavior of the regular solutions
in a characteristic way from asymptotically Minkowskian to that of
deSitter (dS) resp.\ Anti-deSitter space (AdS) depending on the
sign of $\Lambda$. This change is also accompanied by a change of
the moduli space of globally regular (`soliton') and black hole
solutions. While for a given $\Lambda\geq 0$ the generic solution
is singular and there is only a discrete set of soliton solutions
(characterized by the node-number of the YM potential $W$) there
are two types of generic solutions for $\Lambda<0$, the singular
ones and asymptotically AdS solutions with continuously
varying values $W_\infty$ of $W$ at $r=\infty$. This change of the
topology of the moduli space going from negative to positive
values of $\Lambda$ leads to, what was called a `fractal'
structure in \cite{bh00a}.

A further important difference between the dS and AdS case
concerns the stability properties of the solutions. While it was
found that all solutions with $\Lambda>0$ are still unstable like
the Bartnik-McKinnon (BK) solutions \cite{Bart} obtained for
$\Lambda=0$ there exist stable solutions with AdS asymptotics for
negative cosmological constant \cite{bh00a,win99}. In fact, there
seems to be general agreement in the literature
\cite{bh00a,win99,bh00b,hos01,radu} that all solutions with a
nodeless YM potential $W$ are stable. For soliton solutions the
question was studied numerically in \cite{bh00a} and stability was
found for certain solutions with no or one node. An analytical
proof was given in \cite{win99} for black holes with
$W_h>1/\sqrt{3}$ and in \cite{sarbo,sarbe} for nodeless solitons,
however in both cases assuming $\Lambda\ll 0$. How the authors of
\cite{bh00a,win99} arrive at their general statement that all
nodeless solutions are stable remains unclear and in fact we shall 
present convincing arguments that it is wrong. In addition
we give a short discussion of the concept of stability employed in
the above considerations.

As in the case $\Lambda\geq 0$ there are two different types of spherically
symmetric unstable modes, those of  `topological' type
(odd parity) and `gravitational' ones (even parity).
While for the BK solutions there is always an equal number of
unstable modes of both types directly
given by the number of nodes of the solution \cite{lm95} this is not true
for $\Lambda\leq 0$.

Our investigation of the stability properties of the soliton
solutions is based on a general study of the moduli space of the
solutions as a function of $\Lambda$. In particular we find two
discrete sets of lines in the $(\Lambda,b)$ plane, (where $b$
parametrizes the solutions with a regular origin), separating
regions with a different number of unstable modes. First there are
the lines with vanishing $W_\infty$ for solutions with $n$ nodes.
Crossing these lines the number of odd parity unstable modes
changes by one \cite{bh00b,vbls95}. Furthermore there is a
discrete set of lines of zero-modes in the $(\Lambda,b)$ plane
separating regions with a different number of even parity unstable
modes. Besides the region in the moduli space without unstable
modes there are regions with exactly one unstable mode, the
condition for a genuine sphaleron like the traditional one in flat
space \cite{sph}. This may have some interesting consequences for
the application.

Combining numerical with analytical tools we give a detailed description
of the boundary between the domain of asymptotically AdS and the singular
solutions yielding a simple explanation of the `fractal' structure observed
in \cite{bh00a}. In particular, based on earlier work \cite{BFM} on the BK
solutions we will derive a scaling law for the small
$\Lambda$ part of the boundary between these two domains.

For reasons of simplicity we have restricted ourselves to
electrically neutral soliton solutions, but the generalization to
charged solutions and black holes should be straightforward.

\section{Field Equations}\label{chapf}

The action of EYM theory with the cosmological constant $\Lambda$
has the form
\begin{equation} \label{eymlaction}
S_{{\rm EYM}_{\Lambda}}=
\frac{1}{4\pi}\int\Bigl(-\frac{1}{4 G}(R+2\Lambda)
- {1\over 4 g^2}F^{a}_{\mu\nu}F^{a \mu\nu}
\Bigr)\sqrt{-g}d^4x \; ,
\end{equation}
where $G$ is Newton's constant and $g$ is the gauge coupling.
In what follows we put $G=g=1$ choosing suitable units.

We will be interested in spherically symmetric solutions with
regular origin.
For a spherically symmetric space-time the line element decomposes
into
\begin{equation}\label{line}
ds^2=ds^2_2-r^2d\Omega^2\;,
\end{equation}
where $ds^2_2$ is the metric on the 2-d orbit space factorizing out the
action of the rotation group and $d\Omega^2$ the invariant line element of
$S^2$.
The 2-d metric can always be brought to the diagonal form
\begin{equation}\label{2line}
ds_2^2=\Bigl(e^{2\nu}dt^2-e^{2\lambda}dR^2\Bigr)\;,
\end{equation}
where $t$ is a time-like coordinate.
Here we are only considering time-independent solutions and we
naturally choose $t$ to be the Killing time. This means that we can perform
a further dimensional reduction to a 1-d theory.
Then the metric functions
$\nu$, $\lambda$ and $r$
depend only on the radial coordinate $R$.
The latter can be chosen arbitrarily exhibiting $\lambda$ as a gauge
degree of freedom.

For the {\sl SU(2)\/} Yang-Mills field $W_\mu^a$ we use the
standard minimal spherically symmetric (purely `magnetic') ansatz
\begin{equation}\label{Ans}
W_\mu^a T_a dx^\mu=
  W(R) (T_1 d\theta+T_2\sin\theta d\varphi) + T_3 \cos\theta
d\varphi\;,
\end{equation}
where $T_a$ denote the generators of {\sl SU(2)\/}.
The reduced EYM action including a negative cosmological constant $\Lambda$
can be expressed as
\begin{eqnarray}\label{action}
&&
S=-\int dR\,e^{(\nu+\lambda)}
  \Bigl[
  {1\over2}\Bigl(1+e^{-2\lambda}((r')^2
   +\nu'(r^2)')\Bigr)
\nonumber\\&&\qquad\qquad\qquad\qquad\quad
   -e^{-2\lambda}r^2{(W')^2\over r^2}
      -{(1-W^2)^2\over2r^2}+{\lam r^2\over 2}\Bigr]\;.
\end{eqnarray}
Introducing
\begin{equation}\label{first}
N\equiv e^{-\lambda}r'\;,\quad
\kappa\equiv re^{-\lambda}\nu'+N\;,\quad
U\equiv e^{-\lambda}W'\;,
\end{equation}
we write the corresponding Euler-Lagrange equations in first order form
\begin{subeqnarray}\label{feq}
   re^{-\lambda}N'&=&\Bigl(\kappa-N\Bigr)N-2U^2\;,\\
   re^{-\lambda}\kappa'&=&1+2\lam r^2+2U^2-\kappa^2\;,\\
   re^{-\lambda}U'&=&{W(W^2-1)\over r}+\Bigl(N-\kappa\Bigr)U\;,\\
   2\kappa N-N^2&=&2U^2+1-{(1-W^2)^2\over r^2}+\lam r^2\;.
\end{subeqnarray}
The last of these equations is the only remaining diffeomorphism constraint.

We still have to fix the gauge, i.e.\ choose a radial coordinate.
A simple choice is to use the geometrical
radius $r$ for $R$ (Schwarzschild coords.), yielding
$e^{-\lambda}=N$. Putting $\mu\equiv N^2$, $A\equiv e^{\nu}/N$,
and $\Lambda=0$, Eqs.~(\ref{feq}) become equal to Eqs.~(6) in \cite{BFM}.
This coordinate choice has the disadvantage that the equations become
singular at stationary points of $r$. This problem is avoided using
the gauge $e^{\lambda}=r$ introduced in \cite{BFM},
which will also be used here.
In order to stress the dynamical systems character of
Eqs.~(\ref{first},\ref{feq})
we denote the corresponding radial coordinate by $\tau$
and derivatives by a dot.
We thus get
\begin{subeqnarray}\label{taueq}
  \dot r&=&rN\;,\\
  \dot \nu&=&\kappa-N\;,\\
  \dot W&=&rU\;,\\
  \dot N&=&(\kappa-N)N-2U^2\;,\\
  \dot\kappa&=&1+2\lam r^2+2U^2-\kappa^2\;,\\
  \dot U&=&{W(W^2-1)\over r}+(N-\kappa)U\;,
\end{subeqnarray}
Using the constraint Eq.~(\ref{feq}d) the equation for $N$
can also be written as
\begin{equation}\label{Neq}
\dot N={1\over2}\Bigl(1+\lam
             r^2-{(W^2-1)^2\over r^2})\Bigr)-{1\over2}N^2-U^2\;.
\end{equation}

For $\Lambda=0$ these
equations coincide with the Eqs.~(49,50) of \cite{BFM}.
An important quantity is the `mass function'
$m=r(1+\lam r^2/3-N^2)/2$ obeying the
equation
\begin{equation}\label{mass}
\dot m=
 \Biggl(U^2+{(W^2-1)^2\over2r^2}\Biggr)rN\;.
\end{equation}
The value $m_\infty$ of $m$ at $r=\infty$ is the AdS mass of the solution.
Due to the presence of the cosmological constant the vacuum solution
obtained for $W^2=1$ is no longer Minkowski space but Anti-deSitter (AdS)
 space
\begin{equation}\label{AdS}
ds^2=\mu dt^2-{dr²\over\mu}-r^2d\Omega^2\;,
\end{equation}
with
\begin{equation} 
\mu(r)=1+{\lam r^2\over 3}\;,
\end{equation}
while the counterpart of the Schwarzschild solution for $\Lambda<0$ is the
Schwarzschild-AdS solution obtained for
\begin{equation} 
\mu(r)=1-{2M\over r}+{\lam r^2\over 3}\;,
\end{equation}
describing a black hole with AdS asymptotics for $M>0$.

Likewise there is the Reissner-Nord\-str{\o}m-AdS (RN) solution
for $W=0$ and 
\begin{equation}
\mu(r)=1-{2M\over r}+{1\over r^2}+{\lam r^2\over 3}\;,
\end{equation}
carrying an Abelian magnetic charge.
The function $\mu$ has either two simple zeros, a double zero or no zero at
all for positive $r$. The double zero occurs at
$r_0^2=(\sqrt{1+4\lam}-1)/2\lam$ for a certain value $M_0(\lam)$.
As long as $\mu$ has zeros the solution describes a charged
black hole with AdS asymptotics. The zero of $\mu$ at the
smaller value of $r$ corresponds to the Cauchy horizon of the
Reissner-Nord\-str{\o}m black hole.
A double zero of $\mu$ is the position of
an extremal horizon. Solutions without zero have a naked singularity at the
origin.

\section{Global behavior}\label{global}

We want to study the global behavior of solutions with a regular
origin. As in the case without cosmological constant the field
equations (\ref{taueq}) are singular for $r=0$, but using the
methods of \cite{BFM} it is straightforward to show that there is
a 2-parameter family of regular solutions with the behavior
\begin{subeqnarray}\label{rzero}
W(r)&=&1-br^2+\frac{1}{10}b(3b-8b^2+2\lam ) r^4 + O(r^6)\;,\\
\mu(r)&=&1-(4 b^2-\frac{\lam}{3})r^2+\frac{16}{15}b^2(3b+\lam)r^4
        +O(r^6)\;,\\
A(r)&=&A_0\Bigl(1+4 b^2r^2-\frac{4 b^2 }{5}(3b-18b^2+ 2 \lam)r^4
         +O(r^6)\Bigr)\;,
\end{subeqnarray}
where $b$ and $A_0$ are free parameters. A trivial rescaling of
the time coordinate $t$ allows to put $A_0=1$ and we will assume
this condition henceforth for solutions with a regular origin.

We may start with a solution with this behavior at $r=0$ and
integrate the field equations (\ref{taueq}) for increasing $\tau$.
This way we can extend the solution until we hit a singular point
for some finite value $\tau_s$ or otherwise to arbitrarily large
values of $\tau$. As long as $N(\tau)$ is positive $r(\tau)$ will
grow monotonously. However, $N(\tau)$ may have a zero for some
value $\tau_0$. If the function $A$ would stay bounded at $\tau_0$
this would
lead to a zero of the metric coefficient $e^{2\nu}$ implying a
(cosmological) horizon. However, it is easy to see from
Eq.~(\ref{taueq}e) that this is impossible, because $\kappa\geq
1$. Thus the function $\mu$ cannot become negative as is e.g.\
claimed in \cite{bh00a}. In fact it is only the function $N$ that
becomes negative and consequently the function $r(\tau)$ has a
maximum and decreases for larger $\tau$. It follows from
Eq.~(\ref{taueq}d) that $N$, once negative, cannot become positive
again. As in the case without cosmological constant one can show
that $N$ tends to $-\infty$ for some finite value $\tau_s$ and $r$
runs back to zero. There the solution becomes singular in complete
analogy to the case $\Lambda=0$ \cite{BFM}, since the terms
proportional to $\Lambda$ may be neglected for $r\to 0$. One finds
a Reissner-Nord\-str{\o}m type singularity of the metric
\cite{BLM}
\begin{subeqnarray}\label{RN}
  W(r)&=&W_0+{W_0\over 2(1-W_0^2)}r^2+W_3r^3+O(r^4)\;,\\
  \mu(r)&=&{(W_0^2-1)^2\over r^2}-2{M_0\over r}+O(1)\;,\\
 A(r)&=&A_0+O(r^2)\;.
\end{subeqnarray}
with the arbitrary parameters $W_0, W_3, M_0$ and $A_0$.
Topologically the 2-spheres of fixed $r$ of these solutions foliate a 3-sphere
with a singularity at its second `origin', suggesting to call them
`Singular Compact' solutions' (SC).
From the number of parameters it follows that
this is a `generic' class of solutions.

The situation where $N(\tau)$ stays positive is a bit more
involved and we will have to distinguish three cases. Since
$r(\tau)$ is monotonously growing it will either have a finite
limit or it will tend to infinity. In the first case the situation
is practically the same as without cosmological constant.
Following the arguments given in \cite{BFM} one concludes that the
solution exists for all $\tau$ and tends for $\tau\to\infty$ to
the fixed point corresponding to the degenerate horizon of the
extremal RN black hole with
$(r,W,U,N,\kappa)=(r_0,0,0,0,\kappa_0)$ where
$r_0^2=(\sqrt{1+4\lam}-1)/2\lam$ and $\kappa_0^2=\sqrt{1+4\lam}$.
Linearizing the Eqs.~(\ref{taueq}) around this point exhibits the
hyperbolic character of the fixed point with eigenvalues
$\kappa_0$, $-2$, $-\kappa_0(1\pm\gamma)/2$ where
$\gamma^2=1-4/\sqrt{1+4\lam}$. As long as $\lam<15/4$ there are
two complex conjugate eigenvalues leading to damped oscillating
eigenfunctions, while for $\lam\geq 15/4$ these eigenvalues are
real and negative. As in the $\Lambda=0$ case the space-time
splits into two separate parts. The exterior part is the exterior
(to the degenerate horizon) of the extremal RN-AdS black hole and
is geodesically incomplete, while the interior part containing the
origin is a geodesically complete manifold with the topology of
AdS and not a space with a boundary as claimed in \cite{bh00b}.

In the second case $r\to\infty$ there remain two possibilities;
either the solution is asymptotically AdS with finite mass
$M=\lim_{r\to\infty} m(r)$ or it is a new singular solution, with
the mass function $m(r)$ diverging as $r^3$. The AdS type
solutions have the asymptotic behavior (we use again Schwarzschild
coordinates)
\begin{subeqnarray}\label{rinfty}
W(r)&=&W_{\infty}+{ c \over
r}+\frac{3{W_\infty}({W_\infty}^2-1)}{2\lam}\frac{1}{r^2}
+ O\left({1\over r^3}\right) \;, \\
\mu(r)&=&1-{2M\over r}+\frac{\lam}{3} r^2 +( ({W_\infty}^2-1)^2
+\frac{2}{3}\lam c^2)\frac{1}{r^2}+ O\left({1\over
r^3}\right) \;, \\
A(r)&=&A_\infty\Biggl(1 -\frac{c^2}{2r^4}+O\left({1\over
r^5}\right)\Biggr)\;,
\end{subeqnarray}
with the arbitrary parameters $W_\infty$, $c$, $M$, and $A_\infty$.
Like the `SC' solutions this is a generic class of solutions.
From the linear growth of $N=\sqrt{\mu}$ with $r$ it follows that $\tau$
remains finite for $r\to\infty$.

Apart from this regular AdS behavior at infinity there is a singular
class of solutions (called SAdS) with $W$, $U$, $N$, and $\kappa$
growing linearly with $r$.
For their description we introduce new variables
\begin{equation}\label{singads}
\bar W={W\over r}\;,\quad \bar\mu={\mu\over r^2}\;,\quad
{\rm and} \quad V=\bar\mu W'\;.
\end{equation}
In terms of these variables the field equations become
\begin{subeqnarray}\label{nsing}
  r{\bar W}'&=&{V\over \bar\mu}-\bar W\;,\\
  rV'&=&\bar W({\bar W}^2-{1\over r^2})-2V-{2V^3\over {\bar\mu}^2}\;,\\
  r{\bar\mu}'&=&-3\bar\mu-{2V^2\over\bar\mu}-({\bar W}^2-{1\over r^2})^2
   +\lam+{1\over r^2}\;.
\end{subeqnarray}

The singular solution corresponds to a hyperbolic fixed point of
these equations for $r\to\infty$
with finite values $\bar W_0,V_0,\bar\mu_0$ determined by
the vanishing of the r.h.s. These values can be expressed in terms of
$x\equiv{\bar W_0}^2$ obeying the cubic equation
\begin{equation}\label{fixp}
2x^3+4x^2+3x-2\lam(1+x)=0\;,
\end{equation}
which has always one positive root. The values $V_0$ and $\bar\mu_0$ at the
fixed point can be expressed in terms of $x$ as $V_0={1\over 2}x^{3/2}/(1+x)$ and
$\bar\mu_0={1\over 2}x/(1+x)$. Introducing $w=\bar W-\sqrt{x}$,
$v=V-V_0$, and $\chi=\bar\mu-\bar\mu_0$ and linearizing
Eqs.~(\ref{nsing}) in these shifted variables we obtain
\begin{subeqnarray}\label{singlin}
  rw'&=&-w+{2(1+x)\over x}v-{2(1+x)\over\sqrt{x}}\chi\;,\\
  rv'&=&3w-2(1+3x)v+4x\sqrt{x}\chi\;,\\
  r{\chi}'&=&-4x\sqrt{x}w-4\sqrt{x}v+(2x-3)\chi\;.
\end{subeqnarray}
Studying the characteristic equation of this linear system one finds that
its eigenvalues are always real for $x>0$. Since for  $x\to 0$ the eigenvalues
become $1,-3,-4$ and the determinant $16x^3+40x^2+32x+12$ is positive for
$x>0$ there is exactly one unstable mode for any $\lam>0$.

Our numerical analysis shows that any solution with regular origin
belongs to one of the types described above and it would be
desirable to prove this analytically. All the considered solutions
have some well-defined asymptotics and thus there is no indication
of any chaotic behavior.

\section{Moduli space topology}\label{chapnum}

\begin{figure}[p]
\hbox to\linewidth{\hss
    \epsfig{bbllx=56bp,bblly=166bp,bburx=561bp,bbury=565bp,%
    file=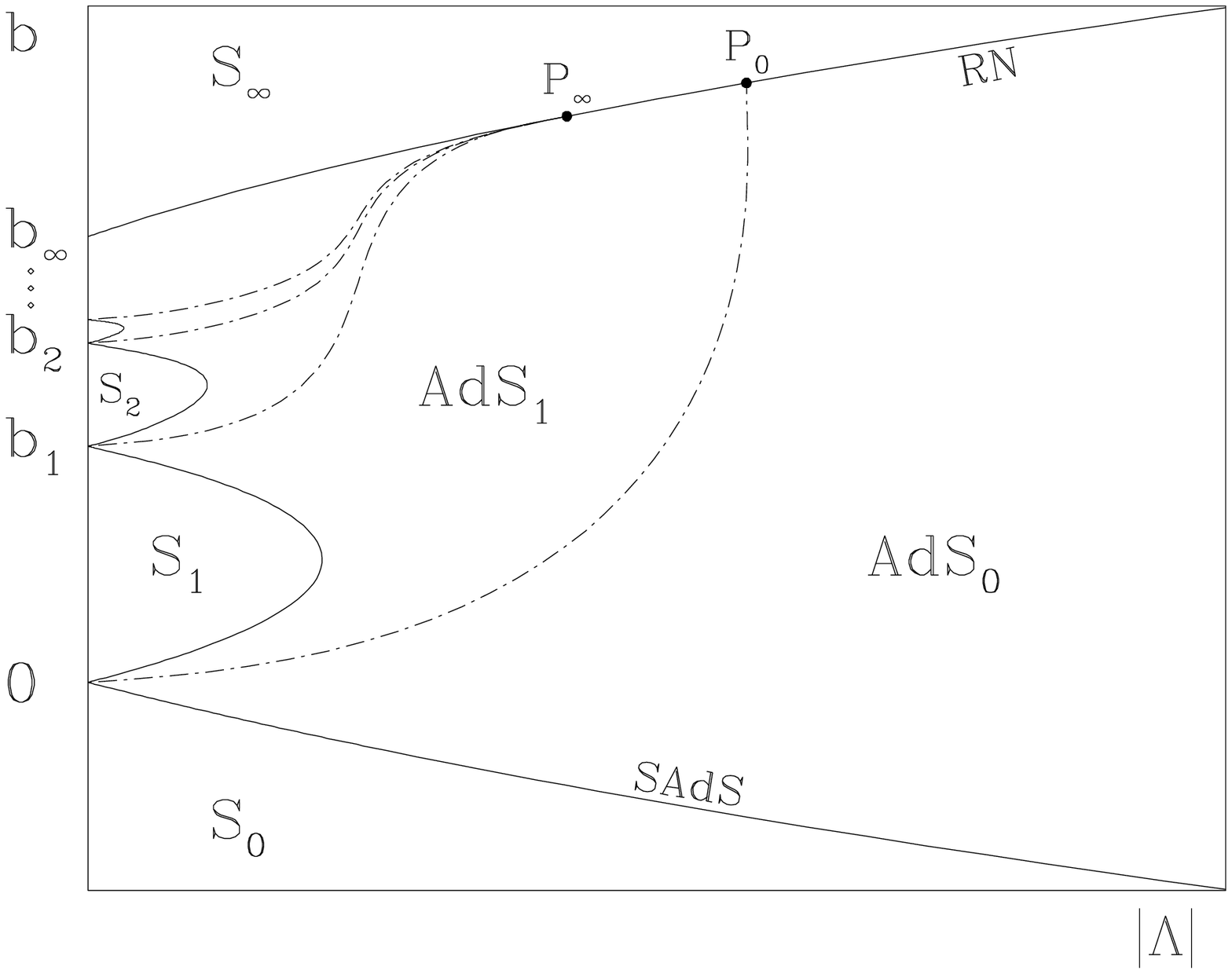,width=0.8\linewidth}\hss}
\caption[figschema]{\label{figschema}%
Schematic plot of the moduli space: the solid lines are the boundary of AdS;
the dashed-dotted lines are curves $W_\infty=0$.}
\end{figure}

Next we want to describe the moduli space of the solutions with regular
origin as obtained from a thorough numerical analysis supported by
analytical arguments.
Our numerical results are obtained using a shooting technique starting from $r=0$
(not from e.g.\ $r=10^{-3}$) combined with integral equations if
necessary as described in \cite{BFMNP}.
For given $\Lambda$ the solutions depend only on the
parameter $b$. In contrast to the case without cosmological constant there
are now two different types of `generic' solutions, the `Singular Compact' ones
running back to $r=0$ already present for $\Lambda=0$ and those with
AdS asymptotics.
Therefore we expect certain open sets in the $(\Lambda,b)$ plane corresponding to
these two types. Fig.~\ref{figschema} gives a schematic description of these
open sets. The domain called `AdS' corresponds to the AdS type solutions.
It has a rather complicated boundary, consisting of the curves `RN' and
`SAdS'. The latter consists of the lower boundary and
infinitely many arcs connecting the points $(\Lambda,b)=(0,b_n)$,
where the $b_n$'s are those of the $n^{\rm th}$ BK solution.%
\footnote
{We include the trivial solution $W\equiv1$ obtained for $b_0=0$.}
The set `AdS' may be further subdivided according to the number of nodes of
$W$ as indicated in Fig.~\ref{figschema}.

The complement of the closure of `AdS' corresponds to the `SC' solutions.
It consists of the disconnected pieces `$S_n$' ($n=0,1\ldots$),
containing solutions with $n$ zeros of $W$ and $|W_0|>1$.
Furthermore there is the set $S_\infty$ containing
solutions with any number of zeros but with $|W_0|<1$. These sets are in close
correspondence with those introduced in \cite{BFM}.
Since the intervals between the $b_n$'s on the $\Lambda$-axis belong to the
sets `$S_n$' and the latter are open, they extend into the region
$\Lambda\ne0$, but as will be explained later, their boundaries
become numerically very quickly
indistinguishable from the $\Lambda$-axis for larger values of $n$.
Varying the parameter $b$ for fixed $\Lambda$ closer and
closer to the axis one intersects more and more of the sets `$S_n$'
alternating with intervals in `AdS'.
This explains the `fractal' structure close to the $\Lambda$-axis observed
in \cite{bh00a}.

The left and lower boundary (`SAdS') of `AdS' corresponds to the singular
solutions with unbounded $W(r)$. Hence $W_\infty$ in `AdS' resp.\ $W_0$
in `$S_n$' becomes arbitrarily large approaching these boundaries.

The upper boundary curve `RN' of `AdS' corresponds to solutions
running into the RN fixed point described in the previous section.
It starts at the value $b_\infty\approx 0.7642$ of \cite{BFM} yielding the
limiting solution of the BK family for $\Lambda=0$. For
$\lam<15/4$ the corresponding solutions have infinitely many zeros
and hence the neighborhood of this piece of the boundary contains
solutions with arbitrarily many zeros.

\begin{figure}[p]
\hbox to\linewidth{\hss
    \epsfig{bbllx=56bp,bblly=166bp,bburx=561bp,bbury=565bp,%
    file=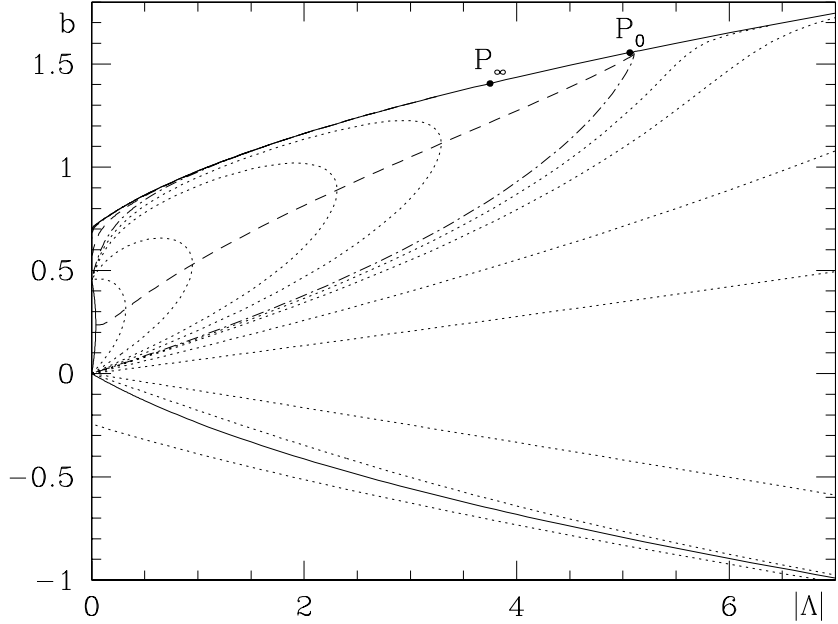,width=0.8\linewidth}\hss}
\caption[figmoduli]{\label{figmoduli}%
The numerically determined moduli space: 
solid and dashed-dotted lines as in Fig.~\ref{figschema},
zero-mode curves are dashed, curves for several $W_\infty$ resp.\
$W_0=$const.\ are dotted.}
\end{figure}

\begin{figure}[p]
\hbox to\linewidth{\hss
    \epsfig{bbllx=56bp,bblly=166bp,bburx=561bp,bbury=565bp,%
    file=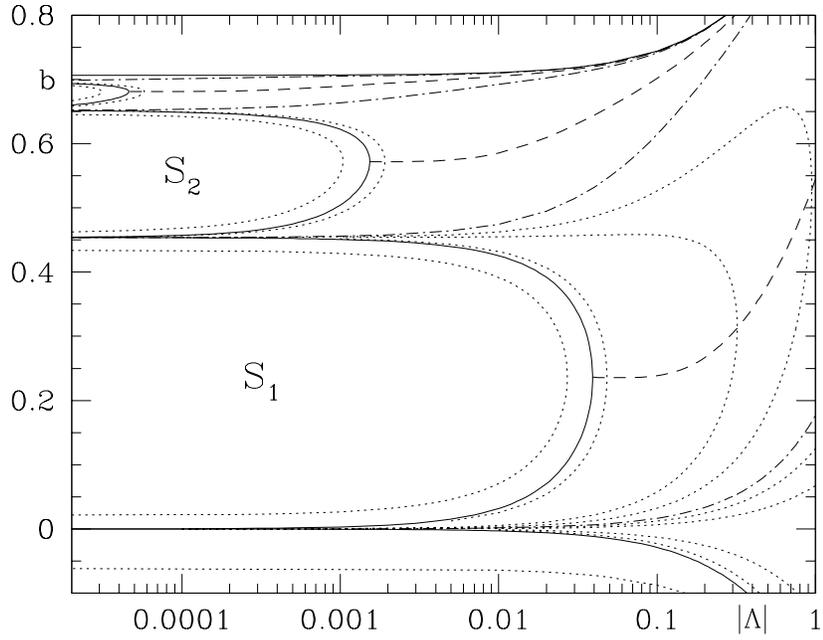,width=0.8\linewidth}\hss}
\caption[figmodlog]{\label{figmodlog}%
Small $\Lambda$ part of Fig.~\ref{figmoduli} with a logarithmic scale.}
\end{figure}

Figs.~\ref{figmoduli} and \ref{figmodlog}
show the same moduli space as Fig.~\ref{figschema}, but
this time with actual numerical data. The dotted curves describe solutions
with constant values $W_\infty$ resp.\ $W_0$.
All the $W_\infty$ curves emanate from
the BK values $(0,b_n)$ on the $\Lambda$-axis. Crossing the dashed-dotted
lines
$W_\infty=0$ the number of zeros of $W$ changes by one. The lowest of these
lines separates solutions without zero from those with exactly one. It
bifurcates with the `RN' boundary curve at the point $P_0$ with
$\lam\approx 5.0646$,
while all the other zero lines bifurcate with `RN' at $P_\infty$ with
$\lam=15/4$.

It is easy to understand that the region where $b\gg\sqrt{\lam}$
belongs to `$S_\infty$'. There the $\Lambda$ term can be neglected
and the situation is the same as for $\Lambda=0$. The same holds
for $b\ll-\sqrt{\lam}$ corresponding to the region $b\ll0$ for
$\Lambda=0$. In the next section we shall derive more detailed
properties of the boundary curves.

\section{Boundary Curves}\label{chapex}

The boundary curves are given by non-generic solutions corresponding to
fixed points with one unstable mode. Hence they have to be determined by
fine-tuning one of the parameters $b$ or $\Lambda$.

In order to determine the `RN' curve one takes solutions running
into the RN fixed point. As mentioned in section \ref{global} such
solutions are oscillating as long as $\lam<15/4$. For $\lam >15/4$
these solutions can have only a finite number of zeros. In fact,
our numerical analysis shows that they have exactly one zero for
$15/4<\lam<L_1$ with $L_1\approx 5.0646$ (the point $P_0$ of
Figs.~\ref{figschema} and~\ref{figmoduli}) and no zeros for larger
values of $\lam$. Therefore all the curves $W_\infty=0$ in `AdS'
with more than one zero merge with `RN' exactly at $\lam=15/4$
(the point $P_\infty$), while the lowest curve with one zero
merges with `RN' at $P_0$. The curves for nodeless solutions with
$0<W_\infty<1$ run out to $\lam=\infty$ eventually approaching the
boundary `RN'.

\begin{figure}[p]
\hbox to\linewidth{\hss
    \epsfig{bbllx=56bp,bblly=166bp,bburx=561bp,bbury=565bp,%
    file=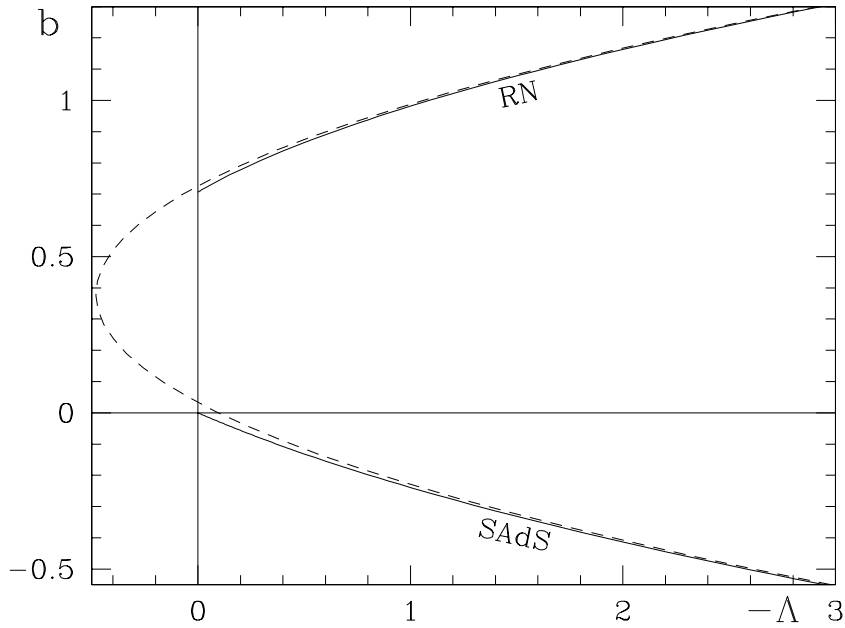,width=0.8\linewidth}\hss}
\caption[figparfit]{\label{figparfit}%
The parabola fit for large $|\Lambda|$ to the boundary curves of AdS.}
\end{figure}

Next we shall consider the asymptotic behavior of the boundary for
large values of $\lam$. As a first step we rescale $r\to \bar
r\equiv\lam^{1/2}r$ in order to remove the explicit $\Lambda$
dependence from the field equations. To retain the expansions
Eq.~(\ref{rzero}b,c) at $r=0$ we have to compensate this by a
corresponding rescaling of $b\to \bar b\equiv\lam^{-1/2}b$. In
fact, the expansion Eq.~(\ref{rzero}a) for $W$ even requires to
put $\bar b=1/2$. Thus asymptotically the boundary of `AdS'
becomes the parabola $\lam=4b^2$. Fig.~\ref{figparfit} shows the
remarkably perfect fit of the boundary with the (empirically
adapted) parabola $\lam=4b^2-3b-0.1$ even for small values of
$\lam$.

The asymptotic form of the solution on `RN' can also be obtained rather simply.
Introducing $\bar W\equiv(W-1)/r$ we find
\begin{subeqnarray}\label{largel}
\bar W&=&-\bar r/2+O(\lam^{-1/2})\;,\\
U&=&-\bar r N+O(\lam^{-1/2})\;.
\end{subeqnarray}
Neglecting non-leading terms the Eq.~(\ref{Neq}) for $N$ becomes
\begin{equation}\label{lleq}
  \dot N=\bar r N{dN\over d\bar r}={1\over2}(1-N^2-2U^2-4\bar W^2+\bar r^2)
    ={1\over2}(1-N^2-2\bar r^2 N^2)\;.
\end{equation}
This linear equation for $N^2=\mu$ can be solved imposing
the boundary condition $\mu(0)=1$ with the simple result
\begin{equation}\label{mueq}
  \mu={{\rm e}^{-\bar r^2}\over \bar r}\int_0^{\bar r}{\rm e}^{r'^2}dr'\;.
\end{equation}
For $\bar r\to\infty$ the solution behaves as
$N=\bar r^{-1}/\sqrt{2}+O(\bar r^{-2})$ and $U=-1/\sqrt{2}+O(\bar r^{-2})$.

It seems to be much more difficult to obtain an asymptotic form of
the SAdS solutions for $\lam\to\infty$ as compared to the
solutions running into the RN fixed point.

The left boundary of `AdS' with its infinitely many arcs
looks at first sight rather complicated. However, making use of the results of
\cite{BFM} we shall derive a simple
scaling law for the asymptotic form of these arcs for large $n$.%
\footnote
{A similar argument was used in \cite{hos01} to derive a scaling law for the
ADM mass of the solutions.}
The basic idea is to relate the neighborhood of the points
$(0,b_n)$ to that of the point $(0,b_0=0)$. As a first step we
consider the solutions with $\lam\ll 1$ and $|b|\ll 1$.

A suitable rescaling in this case is again $\bar
r\equiv\sqrt{\lam}r$. Putting $\bar U\equiv\lam^{-1/2}U$ and
keeping only leading terms in $\lam$ we obtain from
Eqs.~(\ref{taueq})
\begin{subeqnarray}\label{smallb}
  \dot{\bar r}&=&\bar r N\;,\\
  \dot N&=&(\kappa-N)N\;,\\
  \dot\kappa&=&1+2\bar r^2-\kappa^2\;,\\
  \dot W&=&\bar r\bar U\;,\\
  \dot{\bar U}&=&W(W^2-1)/\bar r+(N-\kappa)\bar U\;,
\end{subeqnarray}
together with the constraint
\begin{equation}\label{kap}
2\kappa N-N^2=1+\bar r^2\;.
\end{equation}
Since the equations for $N$ and $\kappa$ are free of matter terms
we obtain the AdS metric. The remaining Eqs.~(\ref{smallb}d,e) for
$W$ and $\bar U$ are just those for the YM field in the AdS
background with $\lam=1$. Clearly we have to rescale also
$b\equiv\lam\bar b$ to obtain the expansion $W=1-\bar b\bar
r^2+O(\bar r^4)$ at the origin.

Numerically one finds a finite interval $\bar b_0<\bar b<\bar b_1$ with
$\bar b_0\approx -0.294582$ and $\bar b_1\approx 2.970497$ for which $W$ stays
bounded. Outside this interval $W$ diverges for finite $\bar r$.
There is an exact solution \cite{Chakra} $W(\bar r)=(1+\bar r^2/3)^{-1/2}$
corresponding to $\bar b=1/6$, which separates solutions with and without
zero, solutions for $\bar b\leq1/6$ having no zero, while solutions with
$\bar b>1/6$ having exactly one.

\begin{figure}[p]
\hbox to\linewidth{\hss
    \epsfig{bbllx=56bp,bblly=166bp,bburx=561bp,bbury=565bp,%
    file=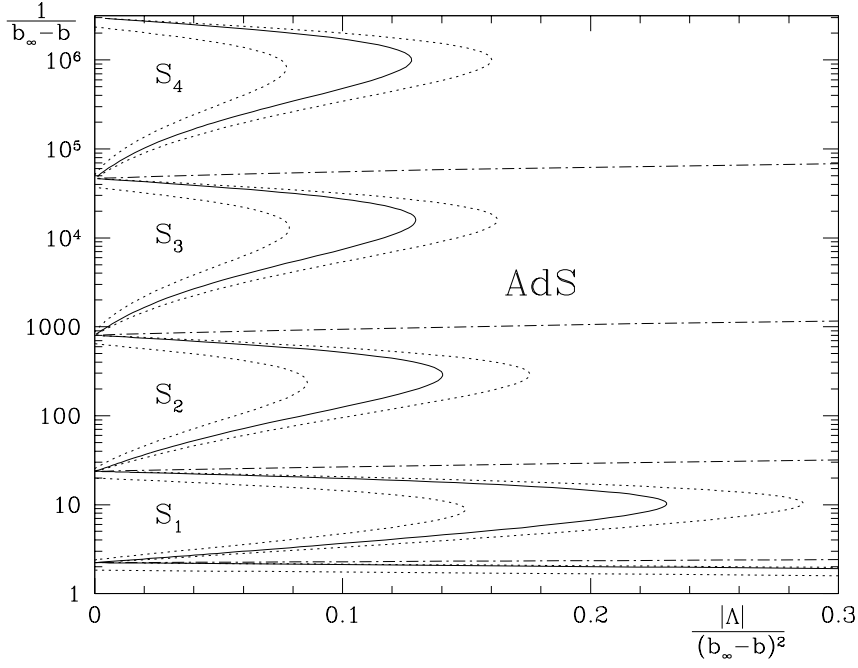,width=0.8\linewidth}\hss}
\caption[figscaled]{\label{figscaled}%
The rescaled regions $S_n$: the dashed-dotted lines are curves
$W_\infty=0$, the dotted lines in AdS are curves for
$|W_\infty|=10$, those in $S_0$, \dots, $S_4$ are curves for
$|W_0|=6.5$, 40, 245, 1505, and 9231.}
\end{figure}

Next we analyze the boundary in the neighborhood of the BK points
$(\Lambda,b)=(0,b_n)$. In \cite{BFM} it was argued that for $b\approx
b_n$ there is some $r_n\approx r_0e^{{n\pi\over\sqrt{3}}}$ such
that for $r\approx r_n$ the solution can be parametrized in the
form
\begin{subeqnarray}\label{rlarge}
N(r)&=&1-{M_n(b,r)\over r}\;,\\
W(r)&=&(-1)^n\Bigl(1-b_n^{(\infty)}(r,b)r^2-{c_n(r,b)\over r}\Bigr)\;,
\end{subeqnarray}
with slowly varying functions $M_n$, $b_n^{(\infty)}$, and $c_n$.
While the coefficients of the convergent modes $M_n(b,r)$ and
$c_n(b_n,r)$ have non-zero limits for $r\to\infty$ that of the
divergent mode $b_n^{(\infty)}(b_n,r)$ vanishes in this limit.
However, assuming transversality its derivative $\partial
b_n^{(\infty)}(b_n,r)/\partial b$ has a non-zero limit
$b_n^{\infty,1}$ and thus $b_n^{(\infty)}\approx
b_n^{\infty,1}(b-b_n)$. Taking $r\gg r_n$ and $b-b_n\ll1$ we can
neglect the terms with $M_n$ and $c_n$ still keeping
$b_n^{(\infty)}r^2\ll1$. This implies the relations $N\approx 1$
and $U/2\approx (W\pm1)/r$ characterizing the stable manifold of
the singular point $r=0$, i.e.\ valid near $r=0$ for solutions
with a regular origin. Thus we succeeded to relate solutions with
$b\approx b_n$ to solutions with $b\approx 0$. All this remains
true for non-zero $\lam\ll1$ and we can repeat the arguments from
above for small $\Lambda$ and $b$. Rescaling again $r\to \bar
r=\lam^{1/2}r$ we will find the same behavior of the solution with
$b\approx b_n$ for $r\gg r_n$ as with $b\approx 0$ for $\bar r>0$
if $\bar b=b_n^{(\infty)}\lam^{-1}\approx
b_n^{\infty,1}(b-b_n)\lam^{-1}$. Arguments analogous to the ones
used in \cite{BFM} to derive scaling laws for the parameters of BK
solutions with large $n$ show that $b_n^{\infty,1}$ scales like
$r_n^{-2}\approx {\rm const.}\cdot e^{-{2n\pi\over\sqrt{3}}}$. Therefore
we expect to find the same linear relation $b=\bar b_{0,1}\lam$
valid for the boundary curves near $(\Lambda,b)=(0,0)$ also near
the points $(0,b_n)$, if we rescale $\Lambda$ by the factor
$e^{-{2n\pi\over\sqrt{3}}}\sim (b_\infty-b)^2$.
Fig.~\ref{figscaled} shows this scaling law is in fact well
satisfied already from $n=3$ on and even holds not only for the
whole arcs joining subsequent $b_n$s, but also the neighboring
curves of constant $W_\infty$ resp.\ suitably rescaled $W_0$.

\begin{figure}[p]
\hbox to\linewidth{\hss
    \epsfig{bbllx=56bp,bblly=166bp,bburx=561bp,bbury=565bp,%
    file=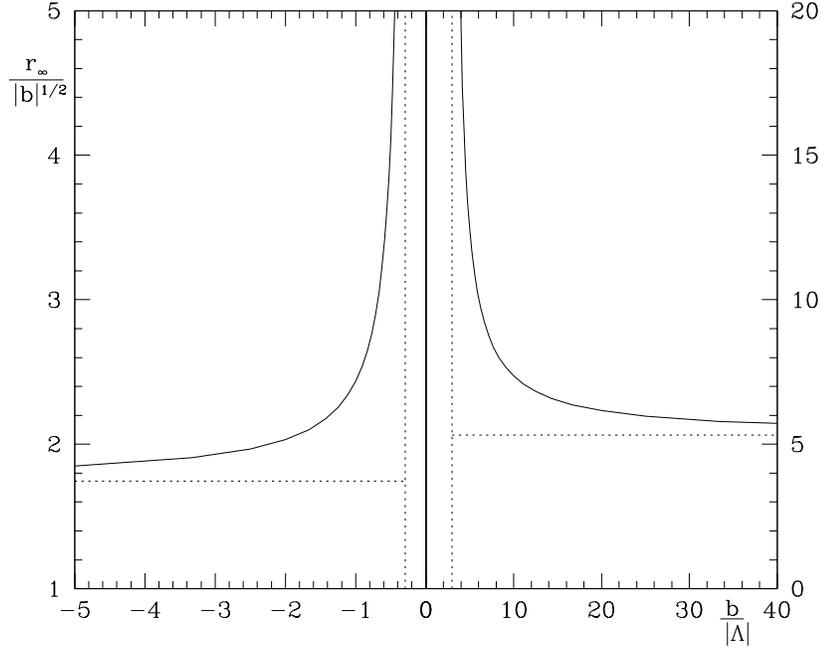,width=0.8\linewidth}\hss}
\caption[figrbarb]{\label{figrbarb}%
The radius $r_\infty$ for singular solutions in the AdS background
as a function of $\bar b=\frac{b}{|\Lambda|}$;
the vertical dotted lines at $\bar b=-0.294582$ and 2.970497
represent the boundary for singular solutions, the horizontal dotted lines
are the BK values $\bar r=1.74575$ and 5.31661 for $\Lambda=0$.}
\end{figure}

The scaling argument for the singular
solutions parametrized by $W_0$ has to be different from that for the AdS ones.
The singular solutions start from a regular
origin $r=0$, have a maximum of $r$ (`equator') at some point $r=r_e$ and
then run back to $r=0$, where $N\to-\infty$.
In \cite{BFM} an asymptotic formula for $r_e$, $W_e$, and $U_e$
for small values of $b$ was derived for the case $\Lambda=0$
\begin{equation}\label{equator}
r_e\approx\bar r|b|^{-{1\over2}}\;,\qquad
|W_e|\approx 2^{-{1\over6}}3^{1\over3}r_e^{2\over3}\;,\qquad
|U_e|\approx {W_e^2\over\sqrt{2}r_e}\;,
\end{equation}
where $\bar r\approx 5.317$ for $b>0$ and $\bar r\approx 1.746$ for $b<0$.
Observe that $W_e$ grows as $|b|^{-1/3}$ for $b\to 0$.
The derivation of these formulae is based on two steps. First the
equation for $W$ is integrated on a flat background, leading to a
divergence of $|W|\approx 1/(\tau_\infty-\tau)$ for some finite value
$\tau_\infty$ and a corresponding finite value
$r_\infty=\bar r|b|^{-1/2}$. The scaling with $|b|^{-1/2}$ is
a consequence of the scale invariance of the flat YM equation.
The two values of $\bar r$ for the two sign choices of $b$ are determined by
a numerical integration of the flat YM equation.
Using this diverging solution for $W$, the equation for $N$ is then integrated from
$N=1$ to $N=0$, the equator, yielding finite values
$\tau_e<\tau_\infty$, $W_e$, $U_e$, and $r_e$. Since this change of $N$
happens on a $\tau$ interval whose length tends to zero for $b\to 0$,
we get  $r_e\approx r_\infty$.
This derivation remains essentially valid
taking into account the cosmological term, but there are two modifications.
Firstly, the function $W$ is now considered in the AdS background. This step
yields functions $\bar r(\bar b)$ (with $\bar b=b/\lam$) for $b>0$ resp.\
$b<0$, replacing the two values of $\bar r$ from the flat case (compare
Fig.~\ref{figrbarb} for a numerical determination).
Secondly, the function $N$ has now to run between the AdS value
$\sqrt{1+\lam r_\infty^2/3}$ and the equator $N=0$. Hence we obtain
\begin{equation}\label{Lequator}
r_e\approx\bar r({b\over\lam})|b|^{-{1\over2}}\;,\quad
|W_e|\approx 2^{-{1\over6}}3^{1\over3}r_e^{2\over3}
      (1+{\lam r_e^2\over3})^{1\over3}\;,\quad
|U_e|\approx {W_e^2\over\sqrt{2}r_e}\;,
\end{equation}

From a systematic point of view it is better not to stop at the
equator as was done in \cite{BFM}, but to continue the solution
all the way back to $r=0$ and to characterize it by its value
$W_0$ at $r=0$. Therefore we decided to plot $W_0$ instead of
$W_e$ (compare Fig.~\ref{figscaled}). For small values of $b$ one
finds $W_0\gg W_e$, since $W_0$ scales like $|b|^{-1/2}$ for $b\to
0$ (remember $W_e\sim|b|^{-1/3}$). In order to describe the growth
of $W$ between the equator and the singular origin we rescale all
the variables $W$ ,$U$, $r$, $N$, $\kappa$ and also the
independent variable $\tau$ by $W\to \bar W|b|^{-1/2}$ etc.\ and
use approximate equations derived from Eqs.~(\ref{taueq}) keeping
only leading terms for $b\to 0$
\begin{subeqnarray}\label{scaletau}
  \dot{\bar r}&=&\bar r\bar N\;,\\
  \dot{\bar W}&=&\bar r\bar U\;,\\
  \dot{\bar N}&=&(\bar\kappa-\bar N)\bar N-2\bar U^2\;,\\
  \dot{\bar\kappa}&=&2\bar U^2-\bar\kappa^2\;,\\
  \dot{\bar U}&=&{\bar W^3\over \bar r}+(\bar N-\bar\kappa)\bar U\;,
\end{subeqnarray}
Observe that there is no more dependence on $\Lambda$ in these
equations. In the limit $b\to 0$ the rescaling stretches the
finite $\tau$ interval to infinite length and the equator becomes
a fixed point of the approximate equations. Because of the
different scaling at the equator all the rescaled variables $\bar
W$ etc.\ vanish at this fixed point like negative powers of $\tau$
with fixed coefficients. The only free parameter (apart from the
sign of $W$) is the value $r_e$ of $r$. Integrating the
Eqs.~(\ref{scaletau}) numerically over the infinite $\tau$
interval from $r=r_e$ to $r=0$ we find $|W_0|\approx 1.14871r_e$.
Using Eq.~(\ref{Lequator}) we obtain the simple relation
\begin{equation}\label{W0}
|W_0|=1.14871{2^{1\over 4}\over
       3^{1\over2}\root4\of{1+{\lam\over3}r_e^2}}|W_e|^{3\over2}\;.
\end{equation}

\begin{figure}[p]
\hbox to\linewidth{\hss
    \epsfig{bbllx=56bp,bblly=166bp,bburx=561bp,bbury=565bp,%
    file=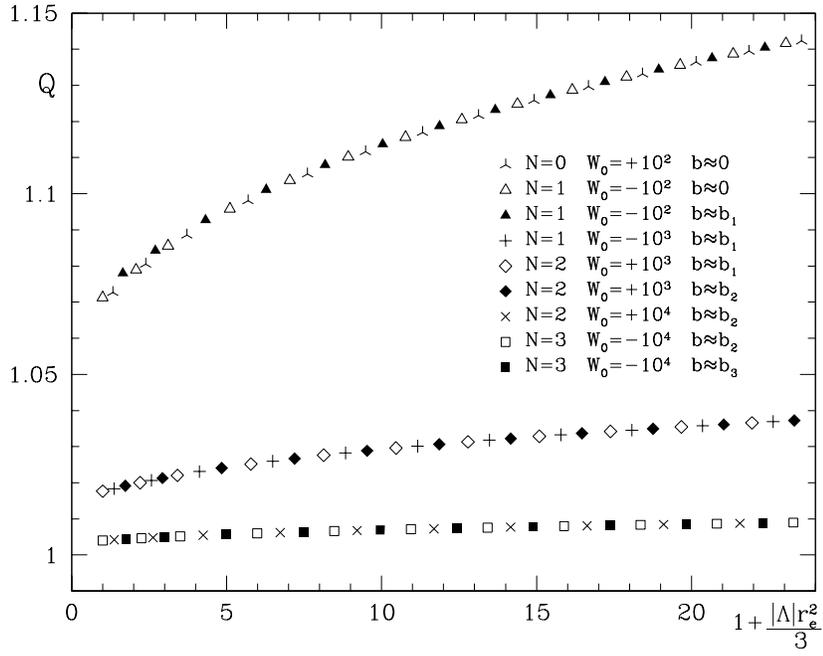,width=0.8\linewidth}\hss}
\caption[figw0wequ]{\label{figw0wequ}%
A plot of the quotient $Q$ of th l.h.s.\ and r.h.s.\ of Eq.~(\ref{W0})
for solutions with $N=0$, \dots, 3 nodes and various values of $W_0$.}
\end{figure}

The argument about the mapping of the neighborhood of the points
$(0,b_n)$ to that of $(0,0)$ by a suitable scaling works as for
the asymptotically AdS solutions with one difference. In contrast
to $W_\infty$ the quantity $W_0$ has to be rescaled, since
$W_0\sim {b_n^{(\infty)}}^{-1/2}\sim e^{n\pi\over\sqrt{3}}$.
Correspondingly the $W_0$ curves in Fig.\ref{figscaled} refer to
$W_0$ values rescaled by a factor $e^{\pi\over\sqrt{3}}\approx 6$
from one $S_n$ domain to the next. Since the relation (\ref{W0})
does not refer to $n$ it should hold equally well in all sectors
$S_n$. Since it is derived for $|W_0|\gg|W_e|\gg1$ it is
easier to test for larger $n$ because of the rescaling.
Fig.\ref{figw0wequ} displays how well this relation is satisfied
for some selected values of $W_0$ and $n$.

\section{Stability}\label{chapstab}

While the BK solutions as well as their counterparts for
$\Lambda>0$ studied in \cite{tmt95,vslhb96,lav97} are known to be
unstable \cite{Strau,bhlsv96} it was observed in
\cite{bh00a,win99,bh00b} that solutions in a certain sub-domain of
`AdS' are stable. Actually, as for BK solutions there are two
different kinds of instabilities. There are instabilities of
`topological' type (`odd parity' sector) comparable to those of
the flat space sphalerons \cite{burz} and `gravitational' ones
(`even parity' sector) \cite{lm95,Strau}. As was shown in
\cite{vbls95} for the BK solutions and generalized to solutions
with $\Lambda>0$ in \cite{bhlsv96}, the number of unstable modes
of the topological type is equal to the number of zeros of $W$.
Although the same seems to be true empirically in the BK case in
the even parity sector there is no general proof (although there
is a rather convincing argument \cite{lm95}). If these rules would
be true also for $\Lambda<0$ one would expect stability in both
sectors for solutions in `AdS' without zero of $W$, as was claimed
in \cite{bh00a}. Although the proof for the odd parity
perturbations can be generalized to $\Lambda<0$
\cite{bh00b,sarbo}, the situation for the even ones is more
subtle. In fact, we shall provide numerical evidence that the
claim in \cite{bh00a,win99,bh00b} is wrong and not all the
nodeless solutions are stable.

Before entering the detailed discussion of our numerical results,
we would like to discuss some more general aspects of the
stability issue. Stability is meant here, as well as in the
literature mentioned above, to be stability on the linearized
level, i.e.\ under infinitesimal perturbations. In the present
context of a stationary background the perturbations are
conveniently decomposed into Fourier modes with respect to the
Killing time $t$ and stability means absence of imaginary
frequencies, i.e.\ exponentially growing modes. The linear
differential equations for these Fourier modes contain in general
some gauge degrees of freedom (independently from the gauge chosen
for the background solution), which have to be fixed. Clearly the
gauge modes have to be separated out and the physical spectrum
should be gauge independent. However, the structure of the
resulting linear system may be quite different and favor some
particular gauge choices. Apart from that there is a choice of
boundary conditions to be made. The latter has two basically
different aspects. On the one hand there may be physical
conditions for the behavior of the perturbations on the boundary,
on the other hand one usually requires self-adjointness of the
linear differential operator. The latter is a technical assumption
made in order to have an expansion theorem for the general class
of perturbations considered. In the case of asymptotically flat
solutions one may restrict the consideration to perturbations of
compact support, although the eigenmodes themselves will generally
not have this property. In this case it is important to have
essential self-adjointness of the linear differential operator on
the perturbations with compact support in order to fix the
spectrum.

Let us now turn to the system under consideration. We shall
restrict ourselves to the spherically symmetric perturbations of
even parity. It is convenient to use the gauge where the
variations are taken at fixed $r$ (`Schwarzschild gauge'), but
observe that this does not require the use of Schwarzschild
coordinates. In fact, it turns out to be convenient to use a
coordinate $\rho$, for which the 2-d metric is conformally flat
putting $dr=A\mu d\rho$. With these conventions the spectral
perturbation problem can be reduced as for $\Lambda=0$ to a
single `Schr\"odinger equation' \cite{win99} for the perturbation
$\delta W$ of $W$
\begin{equation}\label{schroedinger}
\Bigl(-{d^2\over d\rho^2}+U(\rho)\Bigr)\delta W(\rho)=\omega^2\delta
W(\rho)\;,
\end{equation}
with the potential
\begin{equation}\label{potential}
U(\rho)=A^2\mu\Biggl({3W^2-1\over r^2}
        -4{W'^2\over r^2}\Bigl(1-\Lambda r^2-{(W^2-1)^2\over r^2}\Bigr)
        +8W'{W(W^2-1)\over r^3}\Biggr)\;,
\end{equation}
where $W'=dW/dr$.

Next we turn to the boundary conditions. In contrast to the
asymptotically flat case, where $\rho$ runs from 0 to infinity it
runs only over a finite interval $(0,\rho_m]$ in the AdS case.
While the potential $U\approx 2/\rho^2$ at $\rho=0$ is singular it
is easily seen to be finite at $\rho_m$. Correspondingly we are in
the `limit point' case at $\rho=0$, but in the `limit cycle' case
at $\rho_m$ according to Weyl's classification \cite{Hellwig}. This means that
the differential symbol does not define an essentially
self-adjoint operator, but there is a 1-parameter family of
boundary conditions at $\rho_m$. This is related to the fact, that
there are two arbitrary parameters for the background solution at
$r=\infty$ (compare Eq.~(\ref{rinfty})), namely $W_\infty$ and
$c$. The standard choice in the literature is $\delta W_\infty=0$
--- a choice that looks reasonable, although not much more physical
than any condition of the form 
$\cos\alpha\,\delta W_\infty+\sin\alpha\,\delta c=0$. 
Clearly the spectrum of the linear
differential operator depends on the boundary condition. Thus the
class of perturbations vanishing in a neighborhood of $\rho_m$ do
not determine the spectrum uniquely. After all, this is perhaps not
too surprising for a space-time, in which time-like geodesics
reach $r=\infty$ in a finite coordinate time. Anyhow, in order to
be able to compare our results to those in the literature, we make
the choice $\delta W_\infty=0$.

\begin{figure}[p]
\hbox to\linewidth{\hss
    \epsfig{bbllx=56bp,bblly=166bp,bburx=561bp,bbury=565bp,%
    file=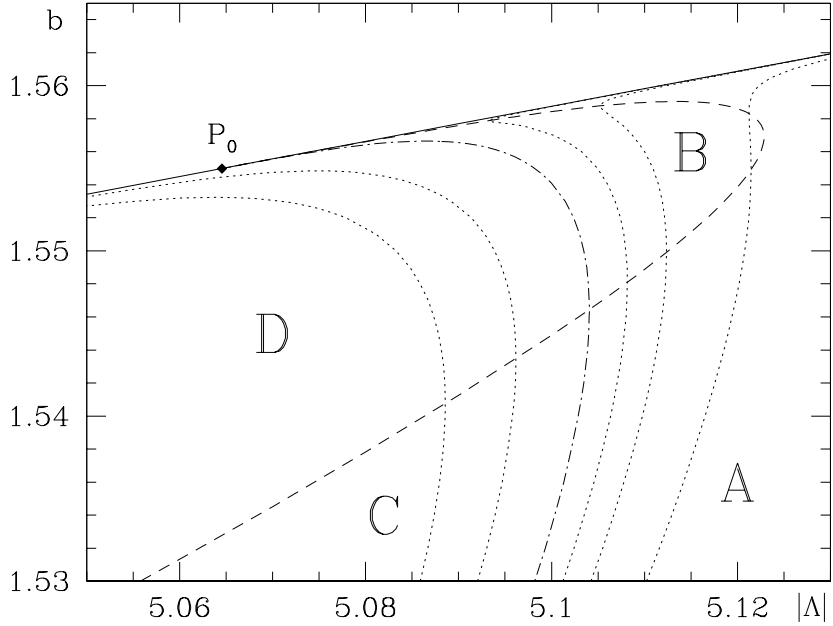,width=0.8\linewidth}\hss}
\caption[figdetail]{\label{figdetail}%
Enlarged view of the moduli space near $P_0$ with $W_\infty$ curves in the
range $\pm 4\cdot10^{-4}$;
the regions A, \dots, D with different numbers of even and odd unstable
modes are separated by the dashed-dotted $W_\infty$ curve and the
dashed zero-mode curve.}
\end{figure}

\begin{figure}[p]
\hbox to\linewidth{\hss
        \epsfig{bbllx=56bp,bblly=166bp,bburx=561bp,bbury=565bp,%
        file=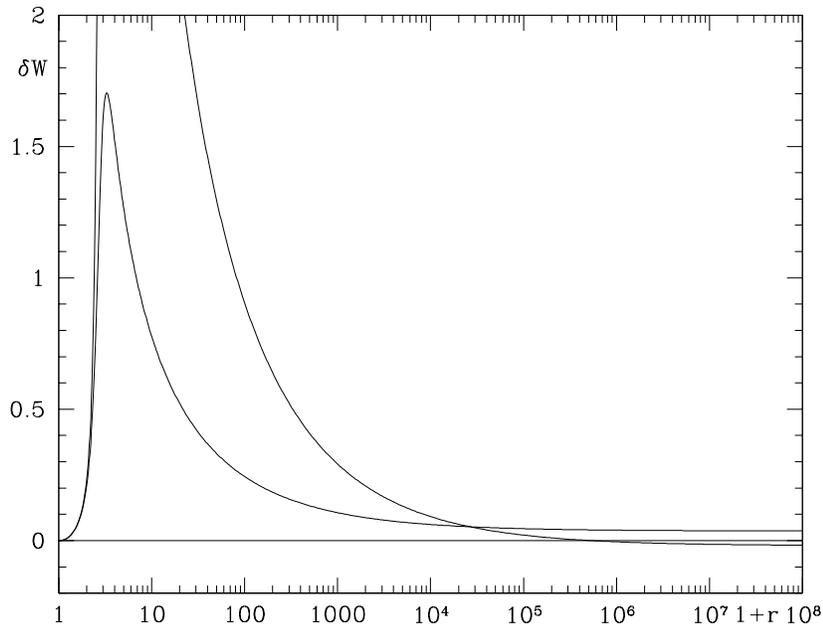,width=0.8\linewidth}\hss}
\caption[figstabil]{\label{figstabil}%
Zero energy wave functions $\delta W$ for
nodeless solutions with $W_\infty=0$ for $b=1.5$,
$\Lambda=-5.075439$ lying below and $b=1.553$, $\Lambda=-5.101469$
lying above the zero-mode curve.}
\end{figure}

A standard way of analyzing a change of stability is to search for
zero modes within a continuous family of solutions \cite{lm95}.
From Fig.~\ref{figmoduli} it can be seen that all the curves with
a fixed value of $W_\infty\ne0$ with at least one node joining the
points $(0,b_n)$ with $(0,b_{n+1})$ have turning points with a
maximal value of $\lam$. These points naturally correspond to
solutions supporting a zero mode. Assuming that the number of
unstable modes changes only at these points we get a relation
between the number of unstable modes of the corresponding BK
solutions and the branches of the $W_\infty$ curves. Of particular
interest are the curves joining $b_0=0$ and $b_1$. As was already
mentioned above, these are just the curves with $W_\infty<0$,
while the $W_\infty=0$ curve bifurcates with the `RN' boundary for
$\lam\approx 5.0646$ and those with $W_\infty>0$ run out to
$\lam=\infty$ approximating the upper resp.\ lower boundary for
$W_\infty<1$ resp.\ $W_\infty>1$. At first sight this seems to
support the hypothesis that the $W_\infty>0$ curves have no
turning points, implying the stability of these solutions. A
careful numerical analysis, however, reveals that this is wrong.
In fact, as Fig.~\ref{figdetail} shows the $W_\infty=0$ curve
still has a turning point at $\lam\approx5.104$. Thus the
solutions on the upper branch of this curve between the turning
point and the bifurcation point are expected to be unstable. A
numerical study of the solution of the linearized equations with
zero eigenvalue supports this result. Fig.~\ref{figstabil} shows
two different `zero-energy wave-functions' $\delta W$ for
solutions with $W_\infty=0$ taken below and above the dashed
stability line in Fig.\ref{figdetail}. The lower one has no zero,
the upper one has one, implying stability resp.\ existence of one
negative eigenvalue corresponding to an unstable mode
\cite{Gelfand}. Neighboring curves with sufficiently small
positive values of $W_\infty<0.0005$ also have a turning point and
follow the $W_\infty=0$ curve close to the `RN' boundary, but then
turn around once more and run out to large values of $\lam$.

Our discussion shows that there is a region in the moduli space where
the solutions are stable. It lies to the right of the lowest $W_\infty=0$
curve and of the curve for the lowest zero mode and is denoted by A
in Fig.~\ref{figdetail}.
Then there are two regions B resp.\ C with solutions that
have exactly one even resp.\ odd
unstable mode, while in D they have one unstable mode of either parity.
Similarly there are further regions with $2n-1$ resp.\ $2n$ unstable modes
between the two types of curves for solutions with $n$ nodes. All these
curves terminate at the point $P_\infty$ on the `RN' boundary.

\section*{Acknowledgements}

G.L. thanks the theory groups of Max-Planck-Institute for Physics
and Max-Planck-Institute for Gravitational Physics for their kind
hospitality during his visits in Munich and Golm,
where part of the work was done.
Work  of G.L. was partly supported by INTAS grant $\#$ 00-00561
and CRDF grant $\#$ 3316.


%

\end{document}